# Unveiling the population of dual- and lensed- AGNs at sub-arcsec separations

F. Mannucci[1], E. Pancino[1,2], F. Belfiore[1], C. Cicone[3], A. Ciurlo[4], G. Cresci[1], E. Lusso[5,1], A. Marasco[1], A. Marconi[5,1], E. Nardini[1], E. Pinna[1], P. Severgnini[6], P. Saracco[6], G. Tozzi[5,1,7], S. Yeh[8]

[1]INAF - Osservatorio Astrofisico di Arcetri, Largo E. Fermi 5, 50125, Firenze, Italy    email: filippo.mannucci@inaf.it
[2]Space Science Data Center – ASI, Via del Politecnico SNC, I-00133 Roma, Italy
[3]Institute of Theoretical Astrophysics, University of Oslo, PO Box 1029, Blindern 0315, Oslo, Norway
[4]Univ. of California Los Angeles, Los Angeles, CA, USA
[5]Dip. di Fisica e Astronomia, Univ. di Firenze, Via G. Sansone 1, 50019, Sesto F. (Firenze), Italy
[6]INAF - Osservatorio Astronomico di Brera, via Brera 28, I-20121, Milano
[7]Cavendish Laboratory, University of Cambridge, 9 JJ Thomson Avenue, Cambridge CB3 0HE, UK
[8]W. M. Keck Observatory, 65-1120 Mamalahoa Highway, Kamuela, HI 96743, USA

**All cosmological models of structure formation predict the existence of a widespread population of dual supermassive black holes in-spiralling inside their common host galaxy, eventually merging and giving rise to intense gravitational waves. These systems can be identified as dual AGNs at kilo parsec separations, but only very few have been confirmed at $z>0.5$. The appearance of multiple AGNs at small angular separations can also be due to gravitational lensing of single AGNs, which are themselves very important systems for many astrophysical topics. Here we present a novel technique, dubbed the Gaia Multipeak (GMP) method, to obtain large and reliable samples of dual/lensed AGN candidates with sub-arcsec separations by looking for AGNs showing multiple peaks in the light profiles observed by the Gaia satellite. All the GMP-selected sources with high resolution images (26 from the HST archive and 5 from dedicated adaptive-optics assisted imaging at the Large Binocular Telescope) show multiple components with sub-arcsec separation pointing toward a very high reliability of the method. By sampling separations down to ~2 kpc at $z>1$, this method allows us to probe the physical processes that drive the inspiralling of a pair of SMBHs inside a single galaxy.**

The multiple supermassive black holes (SMBHs) expected to exist inside many galaxies due to the previous merging events can be revealed by the detection of dual active galactic nuclei (AGNs) separated by up to a few kpc [1, 2]. Identifying these systems is a difficult task [3, 4, 5, 6, 7, 8] and very few confirmed dual AGNs are currently known: only four systems are confirmed with separations below 8 kpc at $z>1$ [9, 10, 11]. This limitation does not allow us to test the predictions of the cosmological models, especially at high redshifts. Lensed AGNs are also of great relevance to many topics of astrophysics and cosmology, such as the measurement of the Hubble constant, understanding of the nature of dark matter, and studying the outflow properties using luminosity boosting [12, 13, 14, 15]. Accurate spectroscopy is needed to distinguish between the two classes of objects [16,17].

The novel selection technique we have developed is based on the all-sky Gaia database and is enabled by the excellent Gaia point-spread function (PSF) in the scan direction (full width at half maximum, FWHM~0.11", [18]). Gaia observations of objects with G>16 mag consist of 1D projection in the along-scan direction of the signal in a 0.71"✕2.1" window [18, 19]. Secondary sources within this window can appear as additional peaks over the light profile of the primary, brighter source.

Starting from the recent data release EDR3 [20], Gaia provides a specific parameter indicating the presence of multiple peaks in these observed 1D light profiles [21, 22, 23], a quantity that can be used to identify multiple sources. Analysing the separation distribution of the apparent pairs in crowded fields, we obtain that our GMP selection can be used to identify dual/lensed AGNs at separations in the range ~0.1"-0.7" (see Methods). By applying the GMP technique to the list of known AGNs with a spectroscopic redshift, we selected 221 systems at redshifts $0.3<z<4$ as dual/lensed AGN candidates (Primary sample, see Extended data). Thirty nine additional sources were identified among the colour-selected AGNs without spectroscopic confirmation (Secondary sample). In total, we have selected 260 multiple AGN candidates. About half of these (119 of 260) correspond to isolated Gaia sources, i.e, systems that are resolved at separations larger than the PSF but are not split into separate entries in the catalogue. To be selected by Gaia, both components must contribute significantly to the Gaia optical G band (~400-950 nm) and, therefore, are unlikely to be heavily affected by dust extinction.

To test the success rate of this selection technique, we have searched the HST archive for images of the selected sample, and found 26 objects. All of these 26 HST images (see Fig. 1) show multiple components at sub-arcsec separations, demonstrating that the GMP method has a very high efficiency in finding compact systems with multiple, point-like components, irrespective of the nature of the companion source.

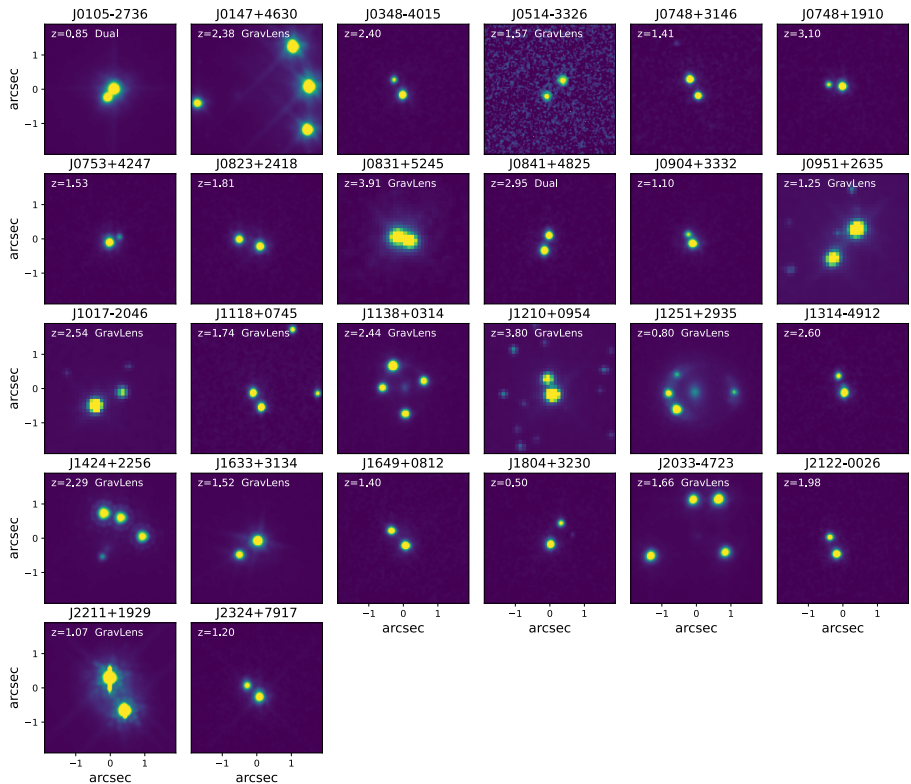

Fig. 1: Archival HST images of the GMP-selected dual/lensed AGN candidates. Different instruments (WFC3, ACS, and WFPC2) and optical filters (from F475W to F814W) were used. All of them show the presence of two or more point sources with sub-arcsec separations. Each image has a field of view of 3.8"✕3.8" and reports the redshift of the source. Previously classified gravitationally lensed and dual systems are labelled as 'GravLens' and 'Dual', respectively

Thirteen out of these 26 objects with HST imaging have been previously classified as gravitationally lensed systems. This population is likely to be significantly overrepresented in this sample because many HST programs have been devoted to the search and observation of lensed systems. One object is a confirmed dual AGN with a very small separation (0.3") [17], while two systems are known alignments between an AGN and a foreground star [24]. The remaining ten systems (see Table 1) lack a previous classification based on well-resolved spectra. Archival HST images of non-lensed AGNs are very rare, and indeed all these ten targets were observed because they had already been selected as dual AGN candidates from the Gaia catalogue through two different methods based on either the astrometric excess noise ("varstrometry") or the presence of several Gaia objects corresponding to one single SDSS target [24] (see Methods). As a consequence, we infer that the sample observed by HST is biased and should not be used to obtain statistical information on the nature of the selected sources.

Tab. 1: Main properties of the ten multiple systems observed by HST and not previously classified. Redshifts for objects 1, 8, and 10 refer to photometric redshifts from [23]. The last four columns report Vega magnitudes.

| # | System | RA         DEC | z | Separation | | Primary | | Secondary | |
|---|--------|----------------|---|------------|---|---------|---|-----------|---|
|   |        |                |   | arcsec | kpc | F475W | F814W | F475W | F814W |
| 1 | J0348-4015 | 03:48:28.67 -40:15:13.2 | 2.4 | 0.50 | 4.2 | 19.62 | 18.87 | 21.43 | 20.63 |
| 2 | J0748+3146 | 07:48:00.55 +31:46:47.4 | 1.408 | 0.53 | 4.5 | 20.49 | 19.71 | 20.90 | 19.25 |
| 3 | J0748+1910 | 07:48:17.13 +19:10:03.06 | 3.096 | 0.40 | 3.1 | 19.29 | 18.45 | 21.28 | 18.50 |
| 4 | J0753+4247 | 07:53:50.58 +42:47:43.9 | 1.528 | 0.32 | 2.7 | 18.42 | 17.60 | 21.85 | 19.10 |
| 5 | J0823+2418 | 08:23:41.08 +24:18:05.6 | 1.814 | 0.63 | 5.3 | 18.55 | 17.21 | 18.94 | 17.53 |
| 6 | J0841+4825 | 08:41:29.77 +48:25:48.5 | 2.948 | 0.46 | 3.8 | 19.84 | 18.71 | 20.17 | 19.28 |
| 7 | J0904+3332 | 09:04:08.67 +33:32:05.27 | 1.103 | 0.30 | 2.5 | 19.01 | 18.42 | 21.05 | 18.75 |
| 8 | J1649+0812 | 16:49:41.30 +08:12:33.5 | 1.4 | 0.61 | 4.9 | 19.43 | 18.22 | 20.09 | 18.93 |
| 9 | J2122-0053 | 21:22:43.01 -00:26:53.8 | 1.975 | 0.52 | 4.4 | 19.56 | 18.63 | 20.74 | 19.15 |
| 10 | J2324+79172 | 23:24:12.70 +79:17:52.4 | 0.4 | 0.49 | 2.7 | 17.94 | 16.22 | 19.29 | 18.24 |

High spatial resolution images of a small sample of 5 GMP targets selected with only the GMP method have been obtained with LBT at near-infrared wavelengths, reaching spatial resolutions of ~0.10" (see Methods). All of them reveal the presence of multiple components, with separations between 0.33" and 0.66", see Fig. 2. Four of these systems (Table 2) have been selected among the sources unsplit in the Gaia catalogue, only identified by the detection of multiple peaks. These observations strengthen the result based on the HST images.

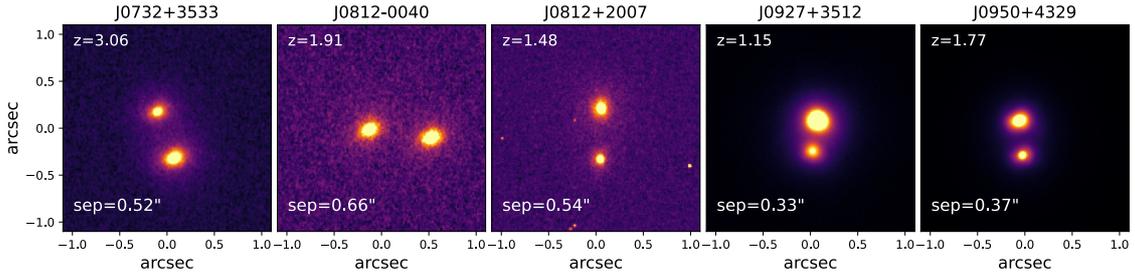

Fig. 2. High resolution, AO-assisted LBT images of 5 GMP-selected systems in the Ks band. Each panel is 2.2"✕2.2" wide. The labels report the redshift of the primary component and the **angular** separation

Tab. 2 : Main properties of the 5 systems observed by LBT. The column "Mult" reports the number of separated Gaia targets within 1.5" from the main AGN, Mult=1 means that the a single source is present in the Gaia archive. The G magnitude of the object J0812+2007 is that of the primary component.

| System | RA           DEC | z | G mag | Mult[1] | Separation | |
|--------|------------------|---|-------|---------|------------|---|
|        |                  |   |       |         | arcsec | kpc |
| J0732+3533 | 07:32:51.57 +35:33:15.34 | 3.06 | 20.31 | 1 | 0.52 | 4.1 |
| J0812-0040 | 08:12:19.34 -00:40:47.97 | 1.91 | 20.36 | 1 | 0.66 | 5.6 |
| J0812+2007 | 08:12:46.41 +20:07:30.18 | 1.48 | 19.11[2] | 2 | 0.54 | 4.6 |
| J0927+3512 | 09:27:48.42 +35:12:41.31 | 1.15 | 15.85 | 1 | 0.33 | 2.7 |
| J0950+4329 | 09:50:31.63 +43:29:08.61 | 1.77 | 17.90 | 1 | 0.37 | 3.2 |

As 119 of the 260 selected targets correspond to single Gaia detections, the GMP method is able to select dual objects that are not split into separate objects by Gaia (see Methods), and is complementary to the techniques sampling larger separations.

One critical point is to understand the nature of the ten unclassified, multiple systems, assessing whether they are chance superpositions with a foreground star, two physically distinct AGNs in the same galaxy, or different gravitational images of the same AGN.

One of the systems with HST imaging, J0841+4825 at z=2.95, has an archival, spatially-resolved HST/STIS optical spectrum (see Fig. 3), sampling the rest-frame wavelengths between 1300Å and 2100Å, obtained as part of the HST GO program 16210 (PI: Xin Liu), selected because both members, with a separation of 0.46", are present as separate entries in the Gaia catalogue. This system was previously classified as a dual AGN even if a ground-based, partially resolved spectrum showed only small differences between the two components [16]. The HST/STIS spectrum has a spatial resolution of ~0.1" and allows us to extract two independent spectra with a signal-to-noise weighted sum along the spatial axis. The fully resolved STIS spectrum reveals the presence of two AGNs with very similar recession velocities, whose difference is consistent with zero, but clear differences in the properties of the emission lines. In particular we detected the NIII]1750 and CIII]1909 emission lines in the fainter source but not in the brighter one. These lines are faint but statistically significant, with signal-to-noise (S/N) ratios of 3.1 and 5.8 for NIII]1750 and CIII]1909, respectively. In case of a lensed system, the spectral differences observed in Fig. 3 could be due to intrinsic variability of the AGN associated with a time delay between the two lensed components. However, the time delay expected for a lensed system at z=2.95 with a 0.5" angular separation is a few days at most [25], while the timescale of variability of the semi-forbidden emission lines in AGN of this luminosity is expected to be a few hundred days in the observer frame [26, 27]. Hence these spectral differences can be more easily attributed to the existence of two distinct AGNs at 3.6 kpc separation, rather than to the time variability of a single, lensed AGN. The non-detection of the lensing galaxy in the HST images [16] further supports the identification of this system as a dual AGN.

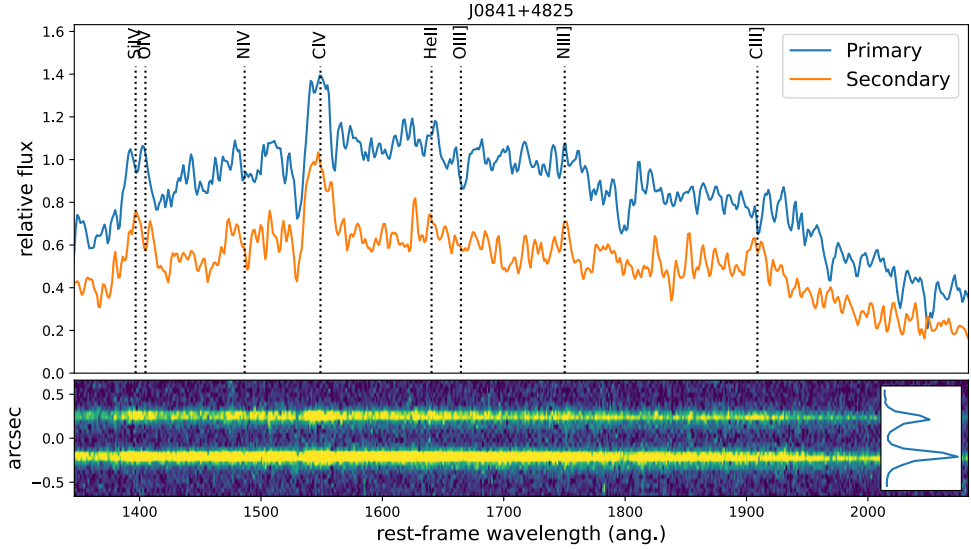

Fig. 3. HST/STIS spectrum of the binary system J0841+4825 at z=2.95. The lower panel shows the original 2D spectra as a function of rest-frame wavelength, and the inset their spatial profile, showing that the two objects are completely spatially resolved at a separation of 0.46". The upper panel shows the extracted, uncalibrated spectra for both objects. The main ultraviolet emission lines expected in AGNs are also shown. The two spectra show different features, in particular the strong CIII]1909 and NIII]1750 lines are only detected in the fainter (secondary) object.

The probability of random alignment with Galactic stars can be robustly estimated in two independent ways: 1) by altering the coordinates of the parent sample and looking for coincidence with other sources of the Gaia catalogue, or 2) by estimating the surface density of Galactic stars down to a given magnitude (see Methods). Both methods are in good agreement with each other and show that contamination from foreground stars is not expected to be a dominant effect, being limited to ~30% of the systems.

The observed colours of the ten previously unclassified systems with multicolour HST images from [24] provide further information about their nature. Fig. 4 shows the observed HST WFC3 F475W-F814W colours as a function of the F475W magnitude for the primary and the secondary components of these ten systems, and compares them with the expected properties of the foreground stars from the Milky Way model TRILEGAL [28]. All the primary members have blue colours, as expected for optically selected AGNs [24], the only exception being J2324+7917, which is not spectroscopically confirmed. The secondary members have a broader colour distribution, extending from blue colours similar to the primaries, to very red colours similar to the field stars. Most of these stars are expected to be of spectral type K and M and have F475W-F814W colours around 2 mag (in the Vega system), with a smaller number of bluer, spectral type G stars. If the companions were stars randomly sampled down to the limiting magnitude, we would expect a colour distribution much more skewed towards red colours. Using the proposed colour separation between AGNs and stars at (F474W-F814W)=1.69 [24], we find that seven out of the ten secondary objects (including J0841+4825, the spectroscopically observed dual AGN in Fig. 3) have blue colours compatible with AGNs, even though two objects are close to the separation line, while the remaining 3 have colours similar to what is expected for stars. The two systems close to the separation line, and even the three systems with red colours, could still be AGNs with various levels of dust extinction, as shown on the right axis of Fig. 4. The number statistics are very low, but our finding is in good agreement with the computed probability of alignments with stars of 30%.

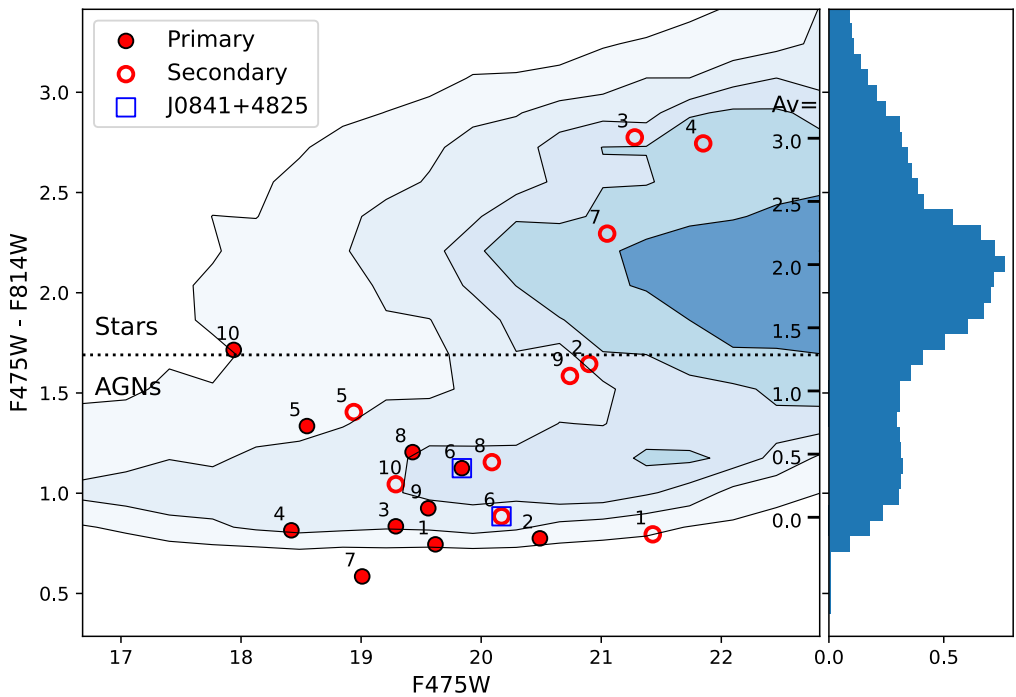

Fig. 4: Optical colours of the primary and secondary components of the spectroscopically confirmed dual AGN J0841+4825 and of the nine unclassified systems observed by HST, numbered as in Table 1. Left: Colour-magnitude diagram. The blue contour plot shows the distribution of the stars from the TRILEGAL model of the Milky Way [28] at galactic latitudes |b|>60deg, the contours enclose 20%, 40%, 60%, 80%, and 90% of the stars from darker to lighter. The dotted line shows the proposed colour separation between stars and AGNs [24]. The scale on the right shows the effect of various levels of dust extinction, computed assuming the extinction law by Calzetti et al [51] and parametrized by $A_V$, on the observed colour for an object at z=1.5 having an intrinsic colour equal to the median of the primary AGNs (F475W-F814W=0.92 mag). Right: the distribution in (F475W-F814W) colour of the model stars.

Additional pieces of information come from the SDSS DR16 spectra [29] available for the majority of the selected systems. These spectra have been obtained using 2" or 3" fibres, therefore they probe the total emission from each system. As both components are detected by Gaia, we expected to see some contribution from both objects in the SDSS spectra. We analysed 100 of these spectra, randomly chosen, by searching for any spectral features that could indicate contamination from a Galactic star. We modelled each SDSS spectrum as the sum of a smooth, power-law continuum and a stellar spectrum at zero radial velocity from the MaStar library [30], fitting for the index of the power-law, the stellar spectrum, and the luminosity ratio between the two. Each best-fitting combination was examined to evaluate the significance of the presence of stellar spectra. Multiple spectral features typical of Galactic stars are detected with high significance in 17% of the spectra, in most cases revealing the presence of an M star. Possible stellar features were detected with lower levels of confidence in 18% more systems. This indicates that between 17% and 35% of the systems have a stellar companion, in agreement with the previous estimates.

Assuming 30% contamination from stars, our GMP method identifies about 0.039% of the AGN parent sample as dual/lensed systems. At the present stage it is difficult to derive robust statistics on the presence of companion AGNs because it is still unknown how the selection function depends on luminosity, separation, and luminosity ratio [31], and what the fraction of lensed systems is. In particular, the luminosity ratio seems to be limited to relatively high values, from 0.04 to 0.74 for dual systems with separate entries in the Gaia catalogue, implying that BHs with similar masses from major mergers are preferentially selected, in agreement with some expectations [2, 32].

Despite the apparently low fraction of selected systems, this technique could already show the presence of a substantial population of multiple systems: assuming that gravitationally lensed systems are a minority, if the duty cycle of the dual AGNs does not depend on the environment and is at the level of ~1% as observed from the clustering properties of the general population [33] and expected by the EAGLE simulations [34], our sample points toward the existence of such similar-mass systems in ~3.9% of the AGNs. The existence of a pair of SMBHs could be revealed in most cases through the detection of an off-centre AGN [e.g. 32, 35, 36]. In contrast, other simulations [32, 37, 38] point towards much longer timescales of activity and higher duty cycles during the late merging stages, thus predicting a much larger fraction of active AGNs. In this case the actual number of dual SMBHs implied by our observation would be much lower.

Despite its success, this method has a few limitations: only AGN pairs in a limited range of separation can be detected (see Methods), complementing other existing methods covering larger separations; the luminosity ratio of the components must be above a certain (still unknown) threshold; it is not sensitive to AGNs with high level of dust absorption; the presence of bright and structured host galaxies is likely to limit the applicability of this method at low redshifts. Physical information, such as the distribution in separation of the dual AGN systems, will be derived from the relative number of objects selected with the same technique.

Nevertheless, this method will open up the possibility of studying the population of dual AGNs inside the same galaxy at z>0.5. This is a crucial observation to understand the processes leading to in-spiralling of the two SMBHs from the moment their host galaxies merge to when they become gravitationally bound. Many authors have performed detailed simulations on the formation of SMBHs through a dual AGN phase and have estimated a number of expected properties of these systems, including luminosity, luminosity ratios, SMBH mass ratio, Eddington ratios for the primary and secondary components, separation distribution, relation to the properties of the host galaxy [2, 32, 34, 36, 37, 38, 39, 40]. By classifying a suitable number of homogeneously selected systems it will be possible to test some of these predictions, and put useful constraints to the models. The absolute number of systems, generally estimated at a few percent of the general AGN population [34, 37], will be more difficult to constrain, as it depends on the total efficiency of the GMP method and on the unknown fraction of low-luminosity and dust-extincted AGNs [e.g., 39] below the detection threshold. In contrast, the distribution of dual AGNs in separation is an observable accessible to our method and is sensitive to a number of physical effects. If the process dominating the reduction in separation of the SMBHs is dynamical friction [41,42], the number of dual AGNs, deconvolved for projection effects [34, 38], is expected to be inversely proportional to their separation. More realistic simulations, considering various dark matter haloes and different galaxy types and dynamics, obtain a broader range of friction characteristics and predict different dependencies with separation [2, 34, 40, 43]. The presence of spiral arms and massive star-forming clumps have also been shown to cause a slowing of the inspiralling at certain distances, further altering the previous predictions [44]. The measurement of the separation distribution is currently only possible for separations above ~10-15 kpc [8], when the SMBHs are not yet part of the same post-merger galaxy, because below this limit a substantial incompleteness is present. The GMP method is expected to cover this gap down to ~2 kpc.

Also, the GMP method is capable of detecting a large number of compact, lensed AGNs to be used in particular to study the 3D structure of the inter- and circum-galactic medium at small separations (e.g., 45, 46, 47), to understand the properties of the AGN hosts [48], and to probe the nature of dark matter by using small substructures present in the lensing galaxy [49], see [50] for a review.

**Methods**

**Detection of multiple peaks and their range of separation**

The Gaia EDR3 catalogue parameter related to the presence of multiple peaks is named ipd_frac_multi_peak, and gives the fraction of Gaia transits/scans with any orientation in which the object appears to have multiple peaks inside the photometric aperture [18, 21, 22]. The GMP method is based on selecting AGNs with high values of this parameter.

The minimum separation that can be sampled by our method is dictated by the Gaia PSF FWHM=0.11" [18, 52]. For each detected source with G<16, Gaia uses a very elongated photometric window with dimensions of about 0.71" ✕ 2.1", centred on the brightest peak. Different scanning directions result in different orientations on the sky of this elongated window. For these objects only a 1D light profile along the scanning direction is saved. In general, secondary objects with a separation larger than about the PSF FWHM but below ~0.35" (half of the smaller size of the window) fall in the same window of the primary, give rise to a single entry in the catalogue, and result in secondary peaks in the 1D light profile. When considering all the scans, these systems are expected to produce large values of ipd_frac_multi_peak. Objects with large separations, above ~1.1", always fall into different windows and thus are separated into different catalogue entries. Objects at intermediate angular distances (between ~0.35" and 1.1") may or may not fall in the same window depending on the angle between the direction of the scan and the separation line between the two sources [53]. Only when they do, a secondary peak can be detected in the 1D light profile, eventually increasing the value of ipd_frac_multi_peak. These objects could result in different behaviours in the final catalogue, including the case in which there are two separate entries, with one or both having a non-zero value of ipd_frac_multi_peak. For this reason the value of this parameter is expected to statistically decrease with separation from ~0.35" to ~1.1".

We can obtain the probability of failing to split two close-by objects into different catalogue entries as a function of their separation by computing the distances among all the sources in small, well-populated fields near the galactic plane, where the apparent pairs are dominated by chance alignment between unrelated objects [22]. The number of pairs increases linearly with separation, therefore the number of unsplit pairs can be estimated by studying the deviation from the linear relationship at small separations. We considered 600 circular regions each 4' in diameter, within 1 deg from the galactic plane, spread across a large range of galactic latitudes, and containing about 500,000 sources. The results for the magnitude range of interest for this project (17<G<20 for the brighter member of the pair, 17<G<21 for the fainter one) are shown in the left panel of Fig. A1. Pairs of stars of these magnitudes with separation below 0.55" are preferentially catalogued as single objects, while above this limit they are usually split. This result implies that most pairs of AGNs that are split by Gaia have separations larger than 0.55", corresponding to ~4.7 kpc at z=2. The right panel of Fig. A1 reports the values of ipd_frac_multi_peak for the main component of a split pair as a function of separation. As expected, the median value decreases with separation. Using a threshold of 10% we can select most of the split pairs up to separations of ~0.7", and some of them up to ~1.2". In conclusion, we expect to detect pairs among both the Gaia unsplit targets with separations of 0.1"−0.6", and split targets with separations of 0.5"−0.7", with a tail up to 1.2".

The future comparison of the observational results with the expectations of the models will require a detailed knowledge of the completeness of the GMP method as a function of separation and luminosity, especially for the unsplit systems. Estimating this completeness is beyond the scope of the present paper, nevertheless we plan to compute this function by studying dense stellar fields observed by HST. These HST images can provide a sample of projected pairs with separations between 0.1" and 0.7" and known magnitudes, and the GMP selection function can be estimated by looking at what fraction of these systems are recovered depending on separation, magnitude, magnitude difference, and, possibly, on other variables. This is possible because the GMP method is not limited to AGNs, but it can be tested and used on any kind of point source.

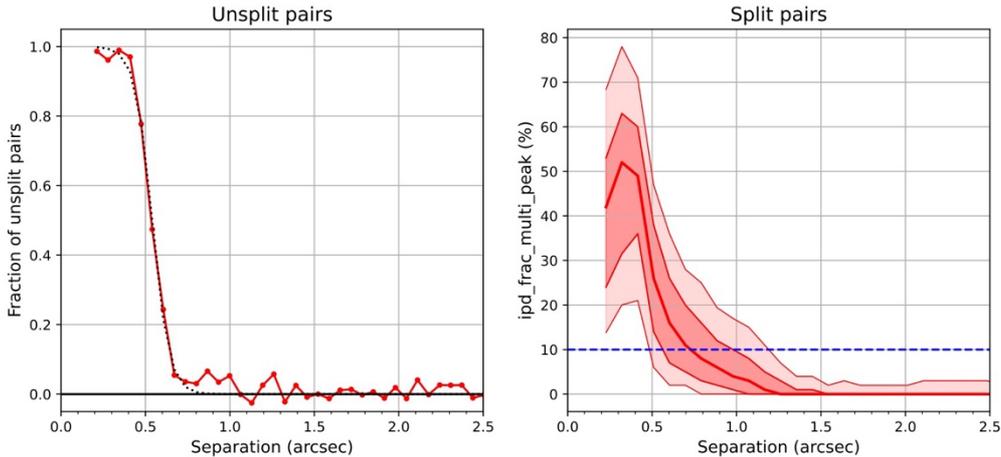

Fig. A1: separation sensitivity of the GMP method. *Left:* Estimated fraction of pairs that are not split in the Gaia EDR3 catalogue as a function of separation. To consider pairs similar to the dual AGNs, we limit the luminosity of the primary and secondary objects to 17<G<20 and 17<G<21, respectively. The dotted line shows an analytic fit to the curve. *Right:* Values of ipd_frac_multi_peak for the primary component of split pairs for the same magnitude ranges as in the left panel. The solid line shows median values as a function of separation, while the shaded

regions show the 25%-75% and 10%−90% distributions. The blue horizontal dashed line shows the threshold used for this work.

**Target selection**

We applied the GMP method to the AGNs of the Milliquas catalogue [54], containing 1.1 million objects selected in different ways, including both spectroscopically- confirmed and colour-selected AGNs with no spectroscopy. We used a cross-matching radius of 1" and tested that the results have no strong dependence on this value. We defined a Primary sample consisting of the objects with: 1) a secure spectroscopic redshift; 2) z>0.3, to avoid AGNs with large and bright host galaxies; 3) G-band magnitude G<20.5, to have reliable values of the Gaia parameters of interest [21, 23]; 4) a large enough distance from the galactic plane (galactic latitude |b|>12°) and from the largest galaxies in the local group (such as LMC and SMC), to avoid most of the contamination from foreground stars. This selection produces about 397,000 AGNs. We searched this catalogue for objects with ipd_frac_multi_peak>10%, obtaining 221 targets (see Fig. A2). We also defined a Secondary sample by relaxing the request for a spectroscopic redshift, but only considering objects with the highest probability of being an AGN (QPCT>=99 in the Milliquas catalogue), obtaining 39 additional candidates out of 39,000 parent objects. This additional sample is likely more prone to contamination by stars, therefore we used the secondary sample only for an additional cross-match with HST, while all the other tests and discussion only refer to the Primary sample.

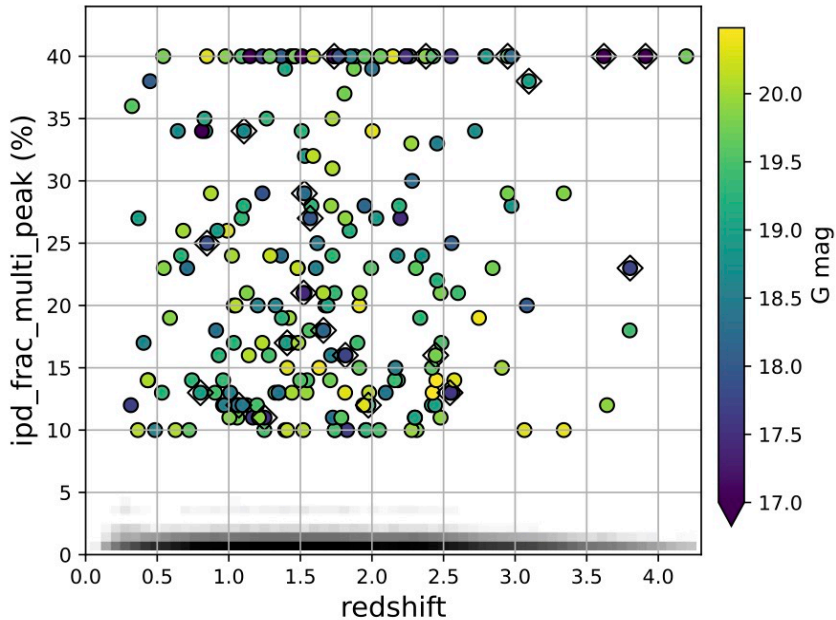

Fig. A2: distribution of ipd_frac_multi_peak of the 221 candidates of the primary sample vs. redshift, colour-coded with their G-band magnitude. For clarity, values of the ipd_frac_multi_peak larger than 40% are plotted at this value. The grayscale image shows the distribution of the input catalogue of AGN, strongly peaked at very low values of ipd_frac_multi_peak. The black diamonds identify the objects with archival HST images.

**Contamination by foreground stars**

Some of the selected systems could be due to the fortuitous alignment of the AGN with a foreground star. Several arguments show that this effect is present, but likely limited to a minority of systems, about 30%.

1. To test the fraction of superpositions with Galactic stars, we randomly altered by a small amount the coordinate of each AGN of the parent population and looked for chance alignment with another point-like object of the Gaia Catalogue [24]. We only considered alignment with fainter objects, otherwise the spectroscopic classification would have been altered. We carried out this computation by adding a random offset of 2 arcmin to sample the actual environment around each object, and looked for objects in the Gaia catalogue that fall within 0.53" of these positions, the median separation of the objects with HST imaging. We found 24 chance superpositions, which corresponds to 30% of the selected sample. This computation has some uncertainties because we do not actually know the selection function of the GMP method as a function of separation, magnitude of the fainter member, luminosity ratio between the two members, and also orientation on the sky due to the non isotropic distribution of the Gaia scan directions. Nevertheless, we expect a chance alignment with a foreground star in a minority of the systems.
2. A similar result is obtained in a more model-dependent way from the expected surface density of Galactic stars and their magnitude from the Milky Way model TRILEGAL [28]. Considering relatively high galactic latitudes, $|b|>60°$, where most of the spectroscopically confirmed AGNs are found, we obtain that 28 systems of the primary parent sample are expected to fall within 0.53" from a star with magnitude F475W<21.3, in agreement with the previous estimate.
3. As explained in the main text, 7 of the 10 unclassified objects with HST images show colours compatible with AGNs (Fig 4).

In conclusion, these three independent tests indicate that about 30% of the multiple systems could actually be due to foreground stars, while the remaining 70% are due to either dual or lensed AGNs. Future spectroscopic observations are needed to determine the nature of each system.

**LBT observations**

The images were obtained at LBT on March 5th, 2022, using the high-resolution LUCI1 camera [55] with a pixel scale of 0.015", assisted by the AO-module the "Single conjugated adaptive Optics Upgrade for LBT" (SOUL) [56, 57]. Natural guide stars (NGS) at separations between 22" and 34" were used for all targets except J0927+3512, which is bright enough to be used to drive the AO module on-axis. In all cases the near-IR Ks-band filter at ~2100 nm was used with an exposure time of 24 min. Data reduction was performed with the custom made pipeline pySNAP. A full analysis of these data will be presented in a future paper.

**Comparison with "varstrometry" and "multiplicity"**

Recently, dual/lensed AGNs candidates were selected using the Gaia archive via two other, conceptually different ways [16, 24, 58]. The first one is based on "varstrometry", namely the extra astrometric jitter due to the uncorrelated luminosity variability of the two components of an unresolved pair. Being based on objects giving rise to single catalogue entries, this method samples a separation range similar to our GMP method. Chen et al [24] find an efficiency of selecting multiple objects of ~53% for targets with spectroscopic redshifts and about 22% for non-spectroscopic ones. Varstrometry only selects dual/lensed AGNs whose variability is substantial [58]. The original target selection for the varstrometric sample was based on the astrometric_excess_noise parameter [22]. In EDR3, the RUWE parameter (renormalized unit weight error) [21] is also provided and is now often used to efficiently identify non-well-behaved sources, using a threshold of 1.4 [23]. High values of RUWE are not required for the GMP selection because large extra jitters are only observed if significant uncorrelated variability is present (see Fig. A3). In fact, also 6 of the 13 gravitationally lensed systems show low RUWE and do not appear in the varstrometric sample. Hwang et al [58] look for AGNs pairs where variability produces non zero values of parallax or proper motion. Our method selects 11 of the 43 objects in their catalogue. Three of the 5 targets of the unbiased sample observed by LBT have low values or RUWE and would not be selected by varstrometry. These results show that the samples selected by the two techniques are for the most part distinct and are likely to have different selection functions.

The second method, referred to as the "multiplicity" selection, identifies AGNs associated with more than one object in the Gaia catalogue with separations up to 3.0" [24, 54]. This method, based on objects that are split by Gaia, samples larger separations, greater than ~0.5", Extended data) and up to 3.0", and is therefore complementary in separation to the GMP method, which is most sensitive below a separation of 0.7". About half (119 out of 260) of the pairs selected by the GMP method, and four of the 5 LBT targets, are not split in the Gaia catalogue, i.e., do not have any companion at separations below 1.5",and would not be selected by the multiplicity method.

All these selection techniques prove that the Gaia catalogue can be efficiently used to identify large numbers of multiple AGNs over a large range of separations and to obtain samples that can finally test one of the central predictions of the current cosmological models

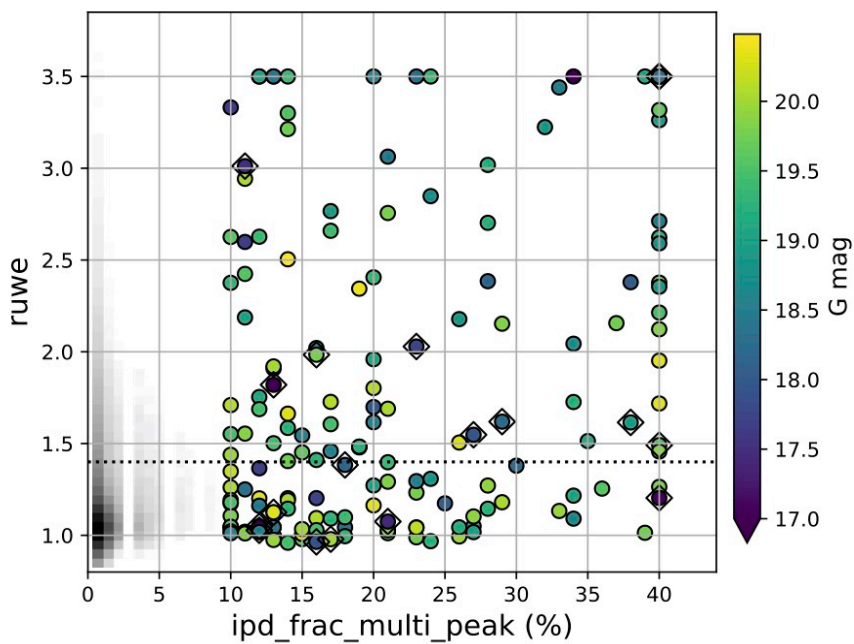

Fig. A3. Values of the ipd_frac_multi_peak (presence of multiple peaks) as a function of RUWE (normalised extra astrometric jitter) for the 221 systems of the primary sample, colour-coded with their G-band magnitude. For clarity, objects with values of RUWE larger than 3.5 and of ipd_frac_multi_peak larger than 40% are plotted at these values, respectively. The grayscale image shows the distribution of the input catalogue of AGN, peaked at low values of ipd_frac_multi_peak but spanning a significant range of RUWE. The dotted line shows the threshold RUWE=1.4. The black diamonds identify the objects with archival HST images

**Data availability**

The Gaia catalogue is publicly available at https://gea.esac.esa.int/ archive/. The HST data are publicly available via the Mikulski Archive for Space Telescopes (MAST) at https://archive.stsci.edu. The Milliquas catalogue of the parent AGNs is available at https://heasarc.gsfc.nasa.gov/W3Browse/all/milliquas.html. SDSS spectra can be downloaded from https://www.sdss.org/dr16.

**Acknowledgements**

This work has made use of data from the European Space Agency (ESA) mission *Gaia* (https://www.cosmos.esa.int/gaia), processed by the *Gaia* Data Processing and Analysis Consortium (DPAC, https://www.cosmos.esa.int/web/gaia/dpac/consortium). Funding for the DPAC has been provided by national institutions, in particular the institutions participating in the Gaia Multilateral Agreement.

This research is based on observations made with the NASA/ESA *Hubble Space Telescope* obtained from the Space Telescope Science Institute, which is operated by the Association of Universities for Research in Astronomy, Inc., under NASA contract NAS 5–26555.

Funding for the Sloan Digital Sky Survey has been provided by the Alfred P. Sloan Foundation, the U.S. Department of Energy Office of Science, and the Participating Institutions. SDSS acknowledges support and resources from the Center for High-Performance Computing at the University of Utah. The SDSS web site is www.sdss.org.

We are grateful to all the LBT staff that performed the requested observations. The LBT is an international collaboration among institutions in the United States, Italy and Germany. LBT Corporation partners are: The University of Arizona on behalf of the Arizona Board of Regents; Istituto Nazionale di Astrofisica, Italy; LBT Beteiligungsgesellschaft, Germany, representing the Max-Planck Society, The Leibniz Institute for Astrophysics Potsdam, and Heidelberg University; The Ohio State University, representing OSU, University of Notre Dame, University of Minnesota and University of Virginia.

**Corresponding author:**

F. Mannucci, filippo.mannucci@inaf.it

**Author Contributions Statement**

F. Mannucci designed and coordinated the work, prepared the figures, and drafted the manuscript. E. Pancino provided the knowledge of the Gaia database parameters. All the authors have contributed to the analysis and interpretation of the data and to the final text.

**Competing Interests Statement**

All the authors have no competing interests